\documentstyle[12pt]{article}

\newcommand{\zp}[3]{Z. Phys.\ C#1 (19#2) #3}
\newcommand{\pl}[3]{Phys.\ Lett.\ #1B (19#2) #3}
\newcommand{\np}[3]{Nucl.\ Phys.\ B#1 (19#2) #3}
\newcommand{\prd}[3]{Phys.\ Rev.\ D#1 (19#2) #3}
\newcommand{\prl}[3]{Phys.\ Rev.\ Lett.\ #1 (19#2) #3}
\newcommand{\sjnp}[3]{Sov.\ J.\ Nucl.\ Phys.\ #1 (19#2) #3}
\newcommand{\jetp}[3]{Sov.\ Phys.\ JETP\ #1 (19#2) #3}

   \def\unlock{\catcode`@=11}

   \unlock

   \def\gsim{\mathrel{\mathpalette\@versim>}}
   \def\lsim{\mathrel{\mathpalette\@versim<}}

   \def\@versim#1#2{\vcenter{\offinterlineskip
        \ialign{$\m@th#1\hfil##\hfil$\crcr#2\crcr\sim\crcr } }}

\begin{document}

\begin{titlepage}



\let\picnaturalsize=N
\def\picsize{1.0in}
\def\picfilename{/home/theory/cschmidt/execs/scipp_tree.eps}

\let\nopictures=Y

\ifx\nopictures Y\else{\ifx\epsfloaded Y\else\input epsf \fi
\let\epsfloaded=Y
{\hbox to\hsize{\hbox{\ifx\picnaturalsize N\epsfxsize \picsize\fi
{\epsfbox{\picfilename}}}\hfill\vbox{


}
}}}\fi


\hspace*{\fill}\parbox[t]{2.8cm}{DESY 93-139 \\ SCIPP 93/35 \\ November 1993}

\vspace*{1cm}

\begin{center}
\large\bf
Dijet Production at Large Rapidity Intervals
\end{center}

\vspace*{0.5cm}

\begin{center}
Vittorio Del Duca \\
Deutsches Elektronen-Synchrotron \\
DESY, D-22607 Hamburg , GERMANY\\
\vspace*{0.5cm}
and\\
\vspace*{0.5cm}
Carl R. Schmidt \footnote{Supported in part by the U.S.
Department of Energy.} \\
Santa Cruz Institute for Particle Physics\\
University of California, Santa Cruz, CA 95064, USA
\end{center}

\vspace*{0.5cm}

\begin{center}
\bf Abstract
\end{center}

\noindent
We examine dijet production at large rapidity intervals at Tevatron
energies, by using the theory of Lipatov and collaborators which resums
the leading powers of the rapidity interval. We analyze the growth of the
Mueller-Navelet $K$-factor in this context and find it to be negligible.
However, we do find a considerable enhancement of jet production at large
transverse momenta.  In addition, we show that the correlation in
transverse momentum and azimuthal angle of the tagging
jets fades away as the rapidity interval is increased.

\end{titlepage}

\section{Introduction}

As the search for the top quark continues, the Tevatron Collider
continues to produce large amounts of data on hadronic jets \cite{CDFD0}.
This data provides a unique opportunity to test our understanding of jet
production at high energies.  The calculation of jet production rates at
hadron colliders is a challenging problem of perturbative QCD, because it
involves many different scales, including:
$\Lambda_{QCD}$, the hadron-hadron center-of-mass energy $\sqrt{s}$, the
parton-parton center-of-mass energy $\sqrt{\hat s}$, and
the momentum transfer $Q$, which is of the order of the transverse
momentum of the jets produced in the hard scattering.

The conventional approach to these calculations is to work at fixed order
in the coupling constant $\alpha_s$, assuming that $\sqrt{s}$, $\sqrt{\hat
s}$, and $Q$ are comparable in size, so that there are no large logarithms
involving them.   The effects of $\Lambda_{QCD}$ are factorized into the
parton structure functions, which are then evaluated at a scale of order $Q$
using the usual DGLAP evolution.  At present the first radiative corrections
to the Born processes are available \cite{ES}.  These yield a more detailed
description of the jet structure, reduce the dependence on the factorization
scale, and are in very good agreement with the data on the one jet inclusive
distribution at large transverse momenta \cite{EKS}.

At the high energies of the Tevatron, however, there may be kinematic
configurations where one cannot ignore the effects of the disparate
energy scales.  In the semihard region, defined as
$s >> Q^2 >> \Lambda_{QCD}^2$, the calculation of jet cross sections is
characterized by the appearance of coefficients containing logarithms of
large ratios of the kinematical invariants.  If no restrictions are made
on $\hat s =x_1\,x_2\,s$, these logarithms will involve the small-$x$
behavior of the structure functions, requiring a more sophisticated
analysis than the usual DGLAP evolution \cite{EC}.  Combined with the
experimental uncertainties in the structure functions at small-$x$, it
appears very difficult to make precise predictions in this kinematic
region.

One way to overcome this problem is to try to disentangle the
different ratios of kinematic invariants in the process.
This can be achieved, for example, by requiring that the parton
momentum fractions, $x_1$ and $x_2$, are large enough that no large
ratios, other than the usual
$Q^2/\Lambda_{QCD}^2$, appear in the evolution of the parton distribution
functions \cite{MN}.  The price to be paid is that the logarithms of
the kind $\hat s/Q^2$ will now appear in the parton
subprocess. These logarithms are of the size of the rapidity interval in the
scattering process.
To realize this configuration experimentally, Mueller and
Navelet proposed to tag two jets at the extremes of the Lego plot in
azimuthal
angle and rapidity, at fixed $x_1$, $x_2$ and transverse momenta
$p_\perp$, and to watch the growth of the dijet inclusive cross section
as the rapidity interval between the $tagging\,jets$ grew with the
center-of-mass energy. To deal with the large logarithms they used the
Balitsky-Fadin-Kuraev-Lipatov theory (BFKL) \cite{BFKL}, which systematically
resums the leading powers in the rapidity interval by using a multigluon
amplitude, with the gluons uniformly filling the rapidity interval
between the tagging jets. The Mueller-Navelet $K$-factor,
defined as the ratio between the resummed and the Born
dijet cross sections at large rapidity intervals and fixed $x$'s,
exhibits the power-like growth in the center-of-mass energy typical of the
BFKL resummation.

In this paper we study the Mueller-Navelet dijet cross section at the
Tevatron energy of $\sqrt{s}$ = 1.8 TeV. Since $\sqrt{s}$ is fixed, we
instead let $x_1$ and $x_2$ vary with the rapidity interval.  At the
same time we retain the Mueller-Navelet requirement that $x_1$, $x_2$ are
large enough that the parton distribution functions can be described by the
DGLAP evolution.  This is done by tagging on the two jets at
the extremes of the rapidity interval $y$ with transverse momenta
larger than some cutoff $p_{\perp\rm min}$.  For reasonable values of $y$
and $p_{\perp\rm min}$ the momentum fractions $x_1, x_2$ will be
sufficiently large.  We can then study the effects of the minijets in
the BFKL resummation as a function of the kinematic variables of the
two tagging jets.

In the exposition of this paper we follow the outline of Ref.~\cite{VC}.
Namely, in sect. 2
we consider the inclusive dijet production $p\,p\rightarrow\, 2\  {\rm
jets}\,+\,X$
at the Born-level, both exact \cite{max} and in the large-rapidity limit.
We then compute the leading logarithmic corrections
at large rapidity as done in Ref.~\cite{MN}. In section 3 we present
numerical results for the inclusive dijet production, with and without
integrating over the jet transverse momenta.  We discuss the effects of the
BFKL resummation on the growth of the Mueller-Navelet K-factor, on
the single jet $p_\perp$ distribution at large $y$,
and on
the correlations of the two jets in transverse momentum and azimuthal angle
as the rapidity interval is increased.  In section 4
we include some
remarks on the range of validity and the limitations of this purely
leading logarithmic calculation, and we present our conclusions.

\section{The Dijet Inclusive Cross section}
We are going to study the semi-inclusive process $p_Ap_B\rightarrow 2\ {\rm
jets}
\,+\, X$ in the semihard regime defined by $\hat s>>Q^2$,
with $Q^2$ being a typical momentum scale in the event, $Q^2\approx
p_{1\perp}\,p_{2\perp}$.  The two tagged jets are chosen with
a large rapidity interval $y=y_1-y_2 \approx
\ln(\hat s/p_{1\perp}p_{2\perp})$.  Other relevant parameters in the
event are the
relative azimuthal angle $\phi$ and the rapidity boost
$\bar y=(y_1+y_2)/2$ of the two jets.

In the semihard, large-$y$ regime we can write the cross section:
\begin{equation}
{d\sigma\over dp_{1\perp}^2 dp_{2\perp}^2 d\phi dy d\bar y}\
=\ \sum_{ij}x_1x_2\,f_{i/A}(x_1,\mu^2)f_{j/B}(x_2,\mu^2)\,
{d\hat\sigma_{ij}\over dp_{1\perp}^2 dp_{2\perp}^2 d\phi}\ ,
\label{general}
\end{equation}
with $f_{i(j)} = Q,\bar Q, G$ labeling the structure function
of the parton species and flavor $i$($j$) = $q,\bar q, g$ inside hadron
$A$($B$). The parton subprocess cross section $d\hat\sigma_{ij}/
dp_{1\perp}^2 dp_{2\perp}^2 d\phi$ contains the sum over all additional
particles ({\it i.e.~minijets}) in the event.  The factorization of the
minijets into the subprocess cross section is possible,
because at large $y$ the initial parton momentum
fractions $x_1$ and $x_2$ are fixed in terms of the two tagged jet
momenta, and are essentially independent of the particles filling the
rapidity interval.  We will arrive at this cross section in several
steps, starting with the exact Born level cross section, taking it to
the $y>>1$ limit, and finally filling in the rapidity interval
with the minijets.

{\it i})\ \underline{Born Level Cross section}.
At the Born level the two partonic jets are produced
back-to-back.  The exact lowest order cross section can be put in the
form (\ref{general}) with the replacement
\begin{equation}
{d\hat\sigma_{ij}\over dp_{1\perp}^2 dp_{2\perp}^2 d\phi}
\ \Rightarrow\ {d\hat\sigma_{ij}\over d\hat
t}\,\delta(p_{1\perp}^2-p_{2\perp}^2)\,\delta(\phi-\pi)\ .
\label{twobody}
\end{equation}
The parton momentum fractions and the subprocess invariants at this
level are given by:
\begin{eqnarray}
x_1\ &=&\ {2 p_\perp e^{\bar y}\over\sqrt{s}}\cosh(y/2) \nonumber\\
x_2\ &=&\ {2 p_\perp e^{-\bar y}\over\sqrt{s}}\cosh(y/2) \nonumber\\
\hat s\ &=&\ x_1x_2s\ =\ 2p_\perp^2 (1+\cosh(y)) \\
\hat t\ &=&\ -p_\perp^2 (1+e^{-y}) \nonumber\\
\hat u\ &=&\ -p_\perp^2 (1+e^{y})\ ,\nonumber\label{invar}
\end{eqnarray}
where $p_\perp=p_{1\perp}=p_{2\perp}$. The subprocess invariants do not depend
on the
rapidity boost $\bar y$. This is a general property, since $\bar y$
parametrizes the collective motion of the parton subprocess in the hadron
reference frame.
The lowest order parton cross sections are well known and can be found
in, for instance, Ref.~\cite{max}.

{\it ii})\ \underline{Large-$y$ Born Cross section}.
We now investigate the lowest order cross section when the rapidity interval
$y$ is large.    For $y>>1$ the lowest order amplitude is
dominated by diagrams with gluon-exchange in the $t$-channel as in
Fig.~1(a).
In this limit the only subprocesses that contribute are $gg\rightarrow gg$
and $qg\rightarrow qg$ and $qq\rightarrow qq$.  We obtain
\begin{equation}
{d\hat\sigma_{gg}\over d\hat t}\ =\ {\pi C_A^2\alpha_s^2
\over 2p_\perp^4}\ ,
\label{largey}
\end{equation}
with $C_A=N_c=3$ the Casimir operator of the adjoint representation.
Similarly, we find
\begin{equation}
 {d\hat\sigma_{qq} \over d\hat t}\ =\
{C_F \over C_A}\, {d\hat\sigma_{qg} \over d\hat
t}\ =\ {C_F^2 \over C_A^2} \, {d\hat\sigma_{gg} \over d\hat t}\ ,
\label{effec}
\end{equation}
with $C_F = (N_c^2-1)/2N_c = 4/3$ the Casimir operator of the
fundamental representation.
Thus, it suffices to consider the subprocess
$gg\rightarrow gg$ and include the other subprocesses by means of the effective
structure function \cite{CM}
\begin{equation}
f_{eff}(x,\mu^2) = G(x,\mu^2) + {C_F\over C_A}\sum_f
[Q_f(x,\mu^2) + \bar Q_f(x,\mu^2)]\ ,
\end{equation}
where the sum is over the quark flavors.
The parton momentum fractions in the large-$y$ limit are
\begin{eqnarray}
x_1\ &=&\ {p_{1\perp}\over\sqrt{s}}e^{(\bar y+y/2)}\ =\
{p_{1\perp}\over\sqrt{s}}e^{y_1}\nonumber\\
x_2\ &=&\ {p_{2\perp}\over\sqrt{s}}e^{(-\bar y+y/2)}\ =\
{p_{2\perp}\over\sqrt{s}}e^{-y_2}\ .
\label{largeyx}
\end{eqnarray}
Equation (\ref{largeyx}) is also valid in the large-$y$ limit when
higher-order
corrections are included, so that $p_{1\perp} \neq p_{2\perp}$.

{\it iii})\ \underline{Minijet-corrected Cross section}.
As discussed in the introduction, going to higher
orders in the coupling constant, i.e.~to multiple parton emission,
we encounter large logarithmic contributions. In the semihard regime,
the BFKL theory \cite{BFKL}\
systematically resums the leading logarithmic terms $\ln(\hat{s}/Q^2)$
by using a multigluon amplitude where the rapidity interval between the
tagging jets is filled with gluons, strongly ordered
in rapidity. This amplitude is shown in Fig.~1(b), where the thick
line represents the resummation of the virtual radiative corrections,
whose effect is to reggeize the gluons exchanged in the $t$ channel.
The real gluons are inserted on these using the Lipatov effective
three-gluon vertex \cite{BFKL}.  The BFKL multigluon amplitude is then
put in a rapidity-ordered phase space, the rapidities of the
gluons are integrated out, and the dependence of the cross section on
the gluon transverse momenta is reduced to the resolution of an
integral equation. Its solution is then convoluted with a jet
emission vertex on each side
of the rapidity interval to give the minijet-corrected parton cross
section for two jets at large-$y$:
\begin{equation}
 {d\hat\sigma_{gg}\over d^2p_{1\perp} d^2p_{2\perp}}\ =\
\biggl[{C_A\alpha_s\over p_{1\perp}^2}\biggr] \,
f(y,p_{1\perp},p_{2\perp}) \,
\biggl[{C_A\alpha_s\over p_{2\perp}^2}\biggr] \ .
\label{cross}
\end{equation}
In this equation
$f(y,p_{1\perp},p_{2\perp})$ is the Laplace transform in the rapidity
interval $y$,
\begin{equation}
 f(y,p_{1\perp},p_{2\perp}) = \int {d\omega \over 2\pi i} e^{\omega y}
f_{\omega}(p_{1\perp},p_{2\perp}),
\label{laplace}
\end{equation}
of the solution of the BFKL integral equation
\begin{equation}
 f_{\omega}(p_{1\perp},p_{2\perp}) = {1 \over (2\pi)^2}
\sum_{n=-\infty}^{\infty}
e^{in(\phi-\pi)} \int_{-\infty}^{\infty} d\nu {(p_{1\perp}^2)^{-1/2+i\nu}
(p_{2\perp}^2)^{-1/2-i\nu} \over \omega-\omega(n,\nu)}.
\label{lip}
\end{equation}
The eigenvalue of the integral equation $\omega(n,\nu)$ is
\begin{equation}
 \omega(n,\nu) = {2 C_A \alpha_s \over\pi} \bigl[ \psi(1) -
 {\rm Re}\,\psi ({|n|+1\over 2} +i\nu) \bigr],
\label{eigen}
\end{equation}
with $\psi$ the logarithmic derivative of the Gamma function.
Substituting (\ref{laplace}) and (\ref{lip}) in (\ref{cross}), and doing the
integral over
$\omega$, the minijet-corrected parton cross section becomes
\begin{equation}
 {d\hat\sigma_{gg}\over dp_{1\perp}^2 dp_{2\perp}^2 d\phi} = {C_A^2\alpha_s^2
\over 4\pi p_{1\perp}^3 \, p_{2\perp}^3}
\sum_n e^{in(\phi-\pi)} \int_0^{\infty} d\nu e^{\omega(n,\nu)\, y}
\cos\left(\nu \, \ln{p_{1\perp}^2 \over p_{2\perp}^2} \right).
\label{mini}
\end{equation}
If we integrate over the azimuthal angle $\phi$ in (\ref{mini}), only the
$n=0$ term survives.

{\it iv})\ \underline{Minijet-corrected Cross section in the Saddle
Point Approximation}.
At very large values of the rapidity interval $y$, the correlations
between the two jets are washed out by the random walk in
transverse momentum space of the gluons exchanged in the $t$ channel.
This can be seen most easily by evaluating (\ref{mini})\ in the saddle-point
approximation. The contribution of (\ref{eigen}) to this equation is
dominated by $n = 0$ and is strongly peaked near $\nu = 0$.
Thus we keep only the first term in the Fourier expansion in $\phi$, and
expand $\omega(\nu) = \omega(0,\nu)$ about $\nu = 0$
\begin{equation}
 \omega(\nu) = A - B\nu^2 +\cdots,
\label{ome}
\end{equation}
with
\begin{equation}
A = {4 C_A \alpha_s \over \pi} \ln2, \quad B = {14 C_A \alpha_s
\over \pi} \zeta(3).
\label{nuzero}
\end{equation}
Then we can evaluate (\ref{mini}) using the saddle-point approximation for
the integral over $\nu$, to obtain
\begin{equation}
 {d\hat\sigma_{gg}\over dp_{1\perp}^2 dp_{2\perp}^2 d\phi} =
{C_A^2 \alpha_s^2 \over 8 p_{1\perp}^3p_{2\perp}^3}
\, {e^{A\, y}\over \sqrt{B \, \pi \, y}} \,
\exp\left(-{\ln^2(p_{1\perp}^2/p_{2\perp}^2) \over 4 B y}\right)\ .
\label{asymp}
\end{equation}
The exponential growth of (\ref{asymp}) with the rapidity interval $y$ is due
to the production of the minijets.

\section{Numerical Results}

We now examine numerically the effects of the minijets at the Tevatron
center-of-mass energy $\sqrt{s}=1.8$ TeV.
We are mainly interested in understanding the behavior of the parton
subprocess, which does not depend on $\bar y$. Therefore,
except where
indicated, we work at fixed $\bar y$ and observe the cross sections
as a function of the rapidity interval $y$.
We chose $\bar y=0$ so that neither $x$ can become too small.
For consistency of notation we will refer to the
leading jet in rapidity as jet 1 and the trailing jet as jet 2  (i.e.,
$y_1=+|y|/2,y_2=-|y|/2$).  Of course, everything is symmetric
under the exchange of the two jets.
We have used the leading order CTEQ structure
functions \cite{cteq} with the renormalization and factorization scale
set to the geometric mean of the transverse momenta of the tagging
jets, $\mu^2=p_{1\perp}p_{2\perp}$.  We shall address some of the difficulties
involved in the choice of scale in the next section.  For $p_{\perp\rm min}
> 10$ GeV and $\bar y = 0$ the parton density functions are always
evaluated at $x> 10^{-2}$, so we are justified in using the DGLAP
evolution in this region of phase space.

We begin by looking for the exponential growth of equation
(\ref{asymp}) in the cross section
as originally suggested by Mueller and Navelet.  To do this we
integrate over the azimuthal angle $\phi$ and over both transverse
momenta above a cutoff of $p_{\perp\rm min} = 20$ GeV.  In Fig.~2(a)
we present this cross section in the first three approximations ({\it
i-iii}) of section 2.  From the plot we see that the large-$y$ Born
cross section is a good approximation to the exact Born level cross
section for large $y\gsim 4$.  However, the minijet-corrected cross
section does not exhibit any great enhancement at large
rapidity.  This is more easily seen in a plot of the K-factor, defined
here as the ratio of the minijet-corrected cross section to the
large-$y$ Born cross section,
\begin{equation}
K\  =\ {d\sigma ({\rm minijet})\over dy d\bar y} \bigg/\,\,
{d\sigma ({\rm large}\!-\!y)\over dy d\bar y}\ .
\label{kfact}
\end{equation}
The $K$-factor is defined so that $K\rightarrow1$ as $y\rightarrow 0$.  In
Fig.~2(b) we
see that the $K$-factor increases until $y\approx 6$, but then quickly
goes to zero \footnote{The fact that $K$ goes below 1 for small $y$
is presumably a computational artifact, arising from the difficulty
in doing the numerical integration over a very sharply peaked function
at small $y$.}.

This effect can be understood if we remember that the rapidity
dependence enters not only
in the BFKL kernel $f(y,p_{1\perp},p_{2\perp})$, but also in
the parton structure functions $f_i(x_1,\mu^2)f_j(x_2,\mu^2)$ where the
momentum fractions are given by (\ref{largeyx}).
The allowed phase space in $p_{1\perp},p_{2\perp}$ is substantially
decreased at large $y$ by the restriction that the momentum fractions
must be less than 1.  The decrease in phase space has a greater effect
on the minijet cross section, with the result that the $K$-factor is
cut off at large $y$.  It requires a much larger range in scales from
$p_{\perp\rm min}$ to $\sqrt{s}$ in order to approach the exponential
growth of the Mueller-Navelet $K$-factor. This is exhibited in
Fig.~3 where we show
the $K$-factor at the Tevatron energy, at a Large Hadron Collider energy
of $\sqrt{s}$~=~15~TeV, and at $\sqrt{s} = 10^5$ TeV.
The full exponential growth is
achieved in the limiting case
of an infinitely large value of $\sqrt{s}$, where the cutoff in the phase
space of the minijet cross section never occurs.

Thus, at the Tevatron energy we must look elsewhere for effects of the
minijets.  In
Fig.~4(a) we show the $p_\perp$ distributions of jet 1 for
a rapidity interval of $y=4$.  We plot the minijet cross section with
two different cutoffs for the second jet, $p_{2\perp\rm min}=10$ GeV and 20
GeV,
while the Born level cross section always has $p_{2\perp}=p_{1\perp}$.
These plots exhibit two effects of the minijets.  We see that the
overall scale of the $p_\perp$ distribution depends strongly on the
minimum $p_\perp$ of the second jet, and that the slope of the
distribution is flatter with a substantial increase at large $p_\perp$.
In Fig.~4(b) at $y=6$ we see an even greater dependence on
$p_{2\perp\rm min}$.  These effects can be
partially understood by the fact that the $p_\perp$ of jet 1 can be
balanced by the smaller $p_\perp$ of jet 2 and the minijets produced in
the rapidity interval.  The lower $x$ values required for this type of
event increase its likelihood relative to
the back-to-back dijet event, which
is all that can occur at the Born level.  It is even possible
for jet 1 to attain transverse momenta that are kinematically
impossible at the Born level.

These arguments suggest that the minijets occuring in the rapidity
interval between the tagged jets will
cause the tagged jets to become uncorrelated.  This can be seen easily
in the minijet formulae by looking at the BFKL kernel as it is
varied from $y=0$ to very large $y$.  For very small rapidities we
approach the Born cross section (\ref{largey}) with
\begin{equation}
f(y,p_{1\perp},p_{2\perp})\ \rightarrow\ \delta(p_{1\perp}^2-p_{2\perp}^2)
\delta(\phi-\pi)\ .
\label{smally}
\end{equation}
The two jets are produced back-to-back in $p_\perp$ and $\phi$.
However, as the rapidity interval becomes large, we have
\begin{equation}
f(y,p_{1\perp},p_{2\perp})\ \rightarrow\ \sim(p_{1\perp}p_{2\perp})^{-1}
\ ,
\label{uncorr}
\end{equation}
and the tagging jets become completely uncorrelated.

The disappearence of correlations as $y$ increases can be seen
dramatically in Fig.~5(a) where we plot the transverse momentum
distribution of jet 1 at a fixed value of $p_{2\perp}=50$ GeV.
For a rapidity interval of $y=2$ the cross section is strongly peaked
near $p_{1\perp}=p_{2\perp}$.  As the rapidity is increased there is a
diffusion of the jet 1 momentum away from the jet 2 momentum until the
peak is practically gone for $y=5$.  In practice jet 2 will be
integrated over some range of transverse momenta, so in Fig.~5(b) we
show the same plot with $p_{2\perp}$ integrated from 50 GeV to 55 GeV
and $|\bar y|\le0.5$. To retain the normalization we have
divided this cross section by 5 GeV.

Similarly, there is also  a reduction of the correlation in the
azimuthal angle $\phi$ as the rapidity interval increases.  This
 can be seen in Fig.~6, where we show the $\phi$
distribution with both jets integrated from $p_{\perp\rm min}=20$ GeV.  The
$\phi$
distribution is normalized to the uncorrelated cross section
$d\sigma/dy\,d\bar y$, so that the area under each curve is equal to 1.
As expected the correlation in $\phi$ decreases as we vary
from $y=5$ to $y=7$.
The decorrelation in $\phi$, however, is slower than the
decorrelation in $p_\perp$, because the
eigenvalue (\ref{eigen}) of the BFKL integral equation is more strongly
peaked in $\nu$ than in $n$.  For example, at $y=5$ the tagging jets
are not correlated any more in $p_\perp$, while they still show a
considerable correlation in $\phi$.

\section{Discussion and Conclusions}

The BFKL analysis that we have been using is a leading logarithmic
approximation.  With this in mind we offer some caveats to our
results and discuss which effects should survive in an exact
calculation.  First, we must state that any of our plots at
very small $y$ are not expected to be very accurate.  For $y\lsim2$
there is even a reasonable discrepancy between the large-$y$ Born Cross
section ({\it ii}) and the exact Born level cross section ({\it i}).
However, we expect that the trends as $y$ is increased should be
apparent even at reasonably small values of the rapidity interval.
In particular the decorrelations in $p_\perp$ and $\phi$ should
definitely increase with $y$.

There are also several ambiguities in our calculation, arising from the
fact that the BFKL analysis assumes little variation in
the $p_\perp$ of the minijets.  For instance, the rapidity variable
used in the standard BFKL analysis is $Y=\ln(\hat s/Q^2)$ where $Q$
is some typical scale of the minijets.  In our calculation we have chosen
$Q^2=p_{1\perp}p_{2\perp}$ so that $Y=y$, the experimental rapidity.
This should not make a significant difference at large $y$,
but it does emphasize the fact that the approximation becomes less
reliable when the transverse momenta of the tagging jets
are not of similar size, and there arise large logarithms of
 order
$\ln(p_{1\perp}/p_{2\perp})$.  In addition, the dependence on jet definition,
cone
size, and other variables are subleading in this analysis at large $y$.

For related reasons the proper renormalization/factorization scale $\mu$
at which the coupling constant is evaluated is not well-determined in
our leading logarithmic analysis.  As required for the BFKL solution, we have
evaluated the coupling for all of the minijets at a single scale of order
$Q$.  Possible choices for $\mu^2$ are $p_{1\perp}p_{2\perp}$ (as we have
used),
$p_{1\perp}^2$, $p_{2\perp}^2$, or ${\rm max}(p_{1\perp}^2,p_{2\perp}^2)$.
At the level of our
approximation, all of these scales are equivalent, but in practice the
choice of scale can make a reasonable difference in the slope of the
$p_\perp$ spectrum \footnote{This ambiguity will arise in any non-exact
calculation involving multiple jets with disparate energies.}.
  However, our main conclusions about the decorrelation
in transverse momentum and azimuthal angle at large $y$, as well as the
increase in the cross section at large $p_\perp$ and $y$ will not change.

Finally, it is interesting to imagine a comparison of our results with
a fixed ${\cal O}(\alpha_s^3)$ calculation.  The fixed order calculation
includes only the effects of up to three parton jets, while the BFKL
resummation includes the leading contributions of amplitudes containing
an arbitrary number of parton jets.  From naive estimates
one might expect that $y\lsim6$ is not large enough to
warrant the use of the full BFKL analysis and that a next-to-leading
order calculation is quite sufficient.  However,
as we have seen, the kinematic phase space is greatly enhanced by
the sharing of transverse momentum among the additional minijets.
This suggests
that it may be necessary to go beyond ${\cal O}(\alpha_s^3)$ at large
rapidity intervals.  Moreover, the BFKL approximation clearly predicts
the main features of the multiple jet emission at large $y$ and readily
suggests the experiments to look for them.  It would be exciting to
compare the predictions of the BFKL resummation given here, as well as
the next-to-leading order calculations, against experiment.

\section*{Acknowledgements}

We wish to thank Bj Bjorken, Jerry Blazey, Terry Heuring, Michael Peskin
and Harry Weerts for useful discussions.

\section*{Figure captions}

\begin{enumerate}
\renewcommand{\labelenumi}{Fig.\arabic{enumi}:}
\item Two jet production amplitude in the large-$y$ limit at
      (a) the Born level and (b) with minijet corrections.

\item (a) Inclusive dijet production at the Tevatron, as a function of
      the rapidity interval $y$. The dashed and dot-dashed lines are
      respectively the exact and large-$y$ Born cross sections, and the solid
      line is the minijet-corrected cross section. (b) The $K$-factor, i.e.
      the ratio of the minijet-corrected cross section to the large-$y$ Born
      cross section, as a function of the rapidity interval $y$.
      The kinematic parameters for both figures are described in the text.

\item $K$-factor as a function of the rapidity interval $y$ at different
      center-of-mass energies. The cutoff on integration
      over both transverse momenta is at $p_{\perp\rm min}$~=~20~GeV. From
      bottom to top,
      the solid lines represent the $K$-factors at Tevatron energies, at
      LHC energies ($\sqrt{s}$=~15~TeV), and at $\sqrt{s} = 10^5$ TeV.

\item $p_\perp$ distribution of jet~1 at (a) $|y|$~=~4 and (b) $|y|$~=~6.
      The dashed and dotdashed lines are respectively the $p_\perp$
      distributions for the exact and the large-$y$ Born cross section,
      for which $p_{1\perp} = p_{2\perp}$. The solid lines are the $p_\perp$
      distributions of jet~1 for the minijet-corrected cross section, with
      two different cutoffs for jet~2, $p_{2\perp\rm min}$~=~10~GeV and 20~GeV.
      Notice that in (b) the dashed and dotdashed lines completely overlap.

\item $p_\perp$ distribution of jet~1 with the jet~2 transverse momentum
      (a) fixed at 50~GeV, and (b) integrated from 50~GeV to 55~GeV. From
      top to bottom, the solid lines are the $p_\perp$
      distributions for the minijet-corrected cross section at $|y|$~=~2,~3,
      4,~5 and 6. In (b) the rapidity boost $\bar y$ is integrated over
      $|\bar y| \le$~0.5.

\item $\phi$ distribution normalized to the uncorrelated cross section
      $d\sigma/dy\,d\bar y$. From top to bottom, relative to the peak, the
      solid lines are the
      $\phi$ distributions for the minijet-corrected cross section at
      $|y|$~=~5,~6 and 7.
\end{enumerate}


\begin{thebibliography}{99}

\bibitem{CDFD0} CDF~Collaboration, preprint Fermilab-Conf-93-204-E,
Fermilab-Conf-93-201-E;\\ D0~Collaboration, preprint Fermilab-Conf-93-050-E,
Fermilab-Conf-93-047-E.

\bibitem{ES} R.K.~Ellis and J.C.~Sexton, \np{269}{86}{445}.

\bibitem{EKS} S.D.~Ellis, Z.~Kunszt and D.E.~Soper, \prd{40}{89}{2188};
\prl{62}{89}{726}; \prl{64}{90}{2121}; \prl{69}{92}{1496}; \prl{69}{92}{3615}.
\\ S.D.~Ellis preprint CERN-TH-6861/93. \\
F.~Aversa, M.~Greco, P.~Chiappetta and J.Ph.~Guillet, \pl{210}{88}{225};
\pl{211}{88}{465}; \np{327}{89}{105}; \zp{46}{90}{253}; \prl{65}{90}{401};
\zp{49}{91}{459}.

\bibitem{EC} J.C.~Collins and R.K.~Ellis, \np{360}{91}{3};\\
S.~Catani, M.~Ciafaloni and F.~Hautmann, \np{366}{91}{135}.

\bibitem{MN} A.H.~Mueller and H.~Navelet, \np{282}{87}{727}.

\bibitem{BFKL}L.N.~Lipatov, \sjnp{23}{76}{338};\\
E.A.~Kuraev, L.N.~Lipatov and V.S.~Fadin, \jetp{44}{76}{443};
\jetp{45}{77}{199};\\ Ya.Ya.~Balitsky and L.N.~Lipatov, \sjnp{28}{78}{822}.

\bibitem{VC} V.~Del~Duca and C.R.~Schmidt, preprint SLAC-PUB-6228, to appear
in Phys. Rev. D.

\bibitem{max} B.L.~Combridge, J.~Kripfganz and J.~Ranft, \pl{70}{77}{234}.

\bibitem{CM} B.L.~Combridge and C.J.~Maxwell, \np{239}{84}{429}.

\bibitem{cteq} J.~Botts et al., \pl{304}{93}{159}.

\end{thebibliography}
\end{document}